\documentstyle[epsfig]{mn}

\newif\ifAMStwofonts
\AMStwofontstrue

\def\beq{\begin{equation}}
\def\eeq{\end{equation}}

\def\spose#1{\hbox to 0pt{#1\hss}}
\def\approxlt{\mathrel{\spose{\lower 3pt\hbox{$\sim$}}
        \raise 2.0pt\hbox{$<$}}}
\def\approxgt{\mathrel{\spose{\lower 3pt\hbox{$\sim$}}
        \raise 2.0pt\hbox{$>$}}}
\def\approxpropto{\mathrel{\spose{\lower 3pt\hbox{$\sim$}}
        \raise 2.0pt\hbox{$\propto$}}}
\mathchardef\twiddle="2218

\def\multleft#1{\hbox to size{\vbox {\halign {\lft{##}\cr #1}}\hfill}\par}
\def\multright#1{\hbox to size{\vbox {\halign {\rt{##}\cr #1}}\hfill}\par}

\def\km{{\rm\thinspace km}}

\def\Msun{\hbox{$\rm\thinspace M_{\odot}$}}

\def\s{{\rm\thinspace s}}
\def\kmps{\hbox{$\km\s^{-1}\,$}}

\def\gtsima{$\; \buildrel > \over \sim \;$}
\def\ltsima{$\; \buildrel < \over \sim \;$}
\def\prosima{$\; \buildrel \propto \over \sim \;$}
\def\gsim{\lower.5ex\hbox{\gtsima}}
\def\lsim{\lower.5ex\hbox{\ltsima}}
\def\simgt{\lower.5ex\hbox{\gtsima}}
\def\simlt{\lower.5ex\hbox{\ltsima}}
\def\simpr{\lower.5ex\hbox{\prosima}}

\title[21cm emission and foregrounds
]{The 21 centimeter emission from the reionization epoch: 
extended and point source foregrounds}         

\author[T. Di Matteo, B. Ciardi \& F. Miniati]{
Tiziana Di Matteo$$, Benedetta Ciardi$$ and Francesco Miniati$$\\  
$$ Max-Planck-Institut f\"ur Astrophysik, Karl-Schwarzschild-Stra\ss e 1, 85748 Garching, Germany\\}

\date{September 2003}

\pagerange{\pageref{firstpage}--\pageref{lastpage}}
\pubyear{2004}
\begin{document}

\maketitle
\label{firstpage}

\begin{abstract}
  Fluctuations in the redshifted 21 centimeter emission from neutral
  hydrogen probe the epoch of reionization. We examine the
  observability of this signal and the impact of extragalactic
  foreground radio sources (both extended and point-like). We use
  cosmological simulations to predict the angular correlation
  functions of intensity fluctuations due to unresolved radio
  galaxies, cluster radio halos and relics and free-free emission from
  the interstellar and intergalactic medium at the frequencies and
  angular scales relevant for the proposed 21cm tomography. In accord
  with previous findings, the brightness temperature fluctuations due
  to foreground sources are much larger than those from the primary
  21cm signal at all scales. In particular, diffuse cluster radio
  emission, which has been previously neglected, provides the most
  significant foreground contamination. However, we show that the
  contribution to the angular fluctuations at scales $\theta \approxgt
  1$ arcmin is dominated by the spatial clustering of bright
  foreground sources. This excess can be removed if sources above flux
  levels $S \approxgt 0.1$ mJy (out to redshifts of $z \sim 1$ and $z
  \sim 2$ for diffuse and point sources respectively) are detected and
  removed. Hence, efficient source removal may be sufficient to allow
  the detection of angular fluctuations in the 21cm emission free of
  extragalactic foregrounds at $\theta \approxgt 1$ arcmin. In
  addition, the removal of sources above $S=0.1$ mJy also
  reduces the foreground fluctuations to roughly the same level as the
  21cm signal at scales $\theta \approxlt 1$ arcmin. This should allow the
  substraction of the foreground components in frequency space,
  making it possible to observe in detail the topology and history of
  reionization.
  \end{abstract}


\section{Introduction}
In the past few years a quantitative study of the high-redshift
intergalactic medium (IGM) and its reionization history has finally
been made possible by the discovery of quasars at $z>5.8$ (e.g. Fan et
al. 2001, 2003). In particular, the detection of a Gunn-Peterson
trough (Gunn \& Peterson 1965) in the Keck (Becker et al. 2001) and
VLT (Pentericci et al. 2002) spectra of the Sloan Digital Sky Survey
quasar SDSS 1030-0524 at $z=6.28$ and in the Keck spectrum of SDSS
1148+5251 at $z=6.37$ (White et al. 2003), has been interpreted as the
signature of the trailing edge of the cosmic reionization epoch. The
recent analysis of the first year of data from the Wilkinson Microwave
Anisotropy Probe (WMAP) satellite on the temperature and polarization
anisotropies of the cosmic microwave background (CMB), infers a mean
optical depth to Thomson scattering $\tau_e \sim 0.17$, suggesting
that the universe was reionized at higher redshift (Kogut et al. 2003;
Spergel et al. 2003). Physically, the CMB and the Gunn-Peterson
trough probe two different stages of reionization, the former being
sensitive to the initial phase, when free electrons appear, the latter
to the residual neutral hydrogen in the latest stages of reionization.
None of the two methods though is able to constrain the exact
ionization level or the details of the reionization history. For this
reason, an alternative way to probe the high-redshift IGM is required.

An optimal experiment for probing the various stages of reionization
is the proposed 21cm tomography. It has indeed been shown that
neutral hydrogen in the intergalactic medium (IGM) and gravitationally
collapsed systems should be directly detectable in emission or
absorption against the cosmic microwave background radiation (CMB) at
frequencies corresponding to the 21cm line (e.g.;Field 1958, 1959; 
Scott \& Rees 1990; Kumar et al.1995). In principle it will
possible to carry out such an experiment with planned high sensitivity
radio telescopes such as the PrimevAl Structure Telescope (PAST)
\footnote{\tt http://astrophysics.phys.cmu.edu}, the Square Kilometer
Array (SKA) \footnote{\tt http://www.nfra.nl/skai} and the LOw
Frequency ARray (LOFAR) \footnote{\tt http://www.astron.nl/lofar}.
The 21cm spectral features will display redshift dependent angular
structure due to evolving inhomogeneities in the gas density field,
hydrogen ionized fraction, and spin temperature.
Several different signatures have been investigated in the recent
literature: the fluctuations in the 21cm line emission induced both by
the inhomogeneities in the gas density and in the ionized hydrogen
fraction (Madau, Meiksin \& Rees 1997; Tozzi et al. 2000; Ciardi \&
Madau 2003, hereafter CM; Furlanetto, Sokasian \& Hernquist 2004) and
by `minihalos' with virial temperatures below $10^4\,$K (Iliev et al.
2002, 2003); the global feature (`reionization step') in the continuum
spectrum of the radio sky that may mark the abrupt overlapping phase
of individual intergalactic HII regions (Shaver et al. 1999); and the
21cm narrow lines generated in absorption against very high redshift
radio sources by the neutral IGM (Carilli, Gnedin \& Owen 2002) and by
intervening minihalos and protogalactic disks (Furlanetto \& Loeb
2002).

While the 21cm tomography proposes to map the topology of the
reionization process and constrain the nature of the ionizing sources,
it remains a challenging project due to foreground contamination from
unresolved extragalactic radio sources (Di Matteo et al. 2002),
free-free emission from the same halos that reionize the universe (Oh
\& Mack 2003) and the Galactic free-free and synchrotron emission
(Shaver et al. 1999). Because the proposed experiments will be carried
out in frequency space as well as angle, recent work has discussed the
possibility of removing the foreground power spectrum components by
comparing maps closely spaced in frequency (Zaldarriaga, Furlanetto \&
Hernquist 2004; Gnedin \& Shaver 2004). The proposed substraction is
nonetheless demanding as the foreground fluctuations overwhelm the
primary 21cm signal by up to three orders of magnitude. In particular,
it will require knowing in detail the spectral behavior and possible
spectral variations of each of the contaminants and their associated
angular power spectra.

In this paper we employ computer simulations of a $\Lambda$CDM
universe to evaluate the foreground brightness temperature
fluctuations due to extragalactic sources that contaminate the 21cm
tomography. In particular, we model the free-free emission from
ionizing sources self-consistently with a viable reionization scenario
and with the associated 21cm IGM emission (as described in CM). Using
the same simulations we adopt simple but physically motivated
prescriptions to model the radio galaxy population; the model
successfully reproduces both the observed luminosity function (at
1.4GHz) and the two point correlation function of the observed radio
sources. Finally, we use a separate simulation (of the same
$\Lambda$CDM universe) that, in addition to dark matter and baryonic
gas, follows the evolution of magnetic fields and cosmic rays, to
estimate the foreground signals due to extended radio sources such as
cluster radio relics and radio halos, which had not been considered
in previous work.

We shall show that the detailed information on the spatial and
redshift distribution of the contaminant sources is crucial for
determining the correct brightness temperature fluctuations due to the
unresolved extragalactic foregrounds on the angular scales
where the primary 21cm signal is expected to peak. By modeling the
foregrounds within the cosmological simulations we are able to study
and map the variation of the angular clustering signal as a function
of the flux cut above which foreground sources will be detected (and
hence will not contribute to the unresolved fluctuating signal). In
particular, we will show that the angular power spectra of unresolved
foreground contaminants become significantly suppressed at scales
$\theta \approxgt 1$ arcmin when sources above flux levels of a
fraction of mJy are removed. 

The structure of the paper is as follows. In Section~2 we describe
the numerical simulations adopted to model the physical processes
producing the extragalactic backgrounds. In Section~3 we briefly
outline the origin of the 21cm emission line from the diffuse IGM used in 
CM, while in Section~4 we show the total comoving emissivities and
spatial correlation functions for the extragalactic foregrounds
modeled within the simulations. In Sections~5, and~6 we calculate
the contribution to the brightness temperature fluctuations to to the
clustering of the different foregrounds and finally, in Section~7 we
give our conclusions.

\section{Numerical Simulations}
In this Section we briefly describe the numerical simulations adopted
to model the various emission processes considered in this paper and
refer to Ciardi, Stoehr \& White (2003, hereafter CSW), Ciardi,
Ferrara \& White, (2003, hereafter CWF) and Miniati (2002) for further
details.

All simulations assume a $\Lambda$CDM ``concordance'' cosmology
with $\Omega_m$=0.3, $\Omega_{\Lambda}$=0.7, $h=$0.7, $\Omega_b$=0.04,
$n$=1 and $\sigma_8$=0.9. These parameters are within the WMAP
experimental error bars (Spergel et al. 2003).  

The simulations used to model the foregrounds produced by radio
galaxies and free-free emission from HII regions are based on the
N-body code {\tt GADGET} (Springel, Yoshida \& White 2001). A
``re-simulation'' technique (Springel et al. 2001, hereafter SWTK) was
employed in order to follow at high resolution the dark matter
distribution within an approximately spherical region with a diameter
of about $50 h^{-1}$~Mpc within a cosmological volume of 479
$h^{-1}$~Mpc on a side (Yoshida, Sheth \& Diaferio 2001).  The
location and mass of dark matter halos was determined with a
friends-of-friends algorithm.  Gravitationally bound substructures
were identified within the halos with the algorithm {\tt SUBFIND}
(SWTK) and were used to build the merging tree for halos and subhalos
following the prescription of SWTK. The smallest resolved halos have
masses of $M \simeq 10^9$~M$_\odot$ (the particle mass is $M_p=1.7
\times 10^8 h^{-1}$~M$_\odot$).  The galaxy population was modeled via
the semi-analytic technique of Kauffmann et al. (1999) and implemented
in the way described by SWTK. We obtain in this way, a catalogue of galaxies
for each of the simulation outputs, containing for each galaxy, among
other quantities, its position, mass and star formation rate. We used
the galaxies within the $50 h^{-1}$~Mpc region, coupled with the
prescriptions described in Section~4, to model the redshift
distribution and luminosity function of radio galaxy population and
the free-free emission from the interstellar medium (ISM).

Within the high resolution spherical subregion, a cube of comoving
side $L=20 h^{-1}$~Mpc was used to study the details of the
reionization process, using the radiative transfer code {\tt CRASH}
(Ciardi et al. 2001; Maselli, Ferrara \& Ciardi 2003) to model the
propagation of ionizing photons into the IGM. Several sets
of radiative transfer simulations were run in CSW and
CFW, with different
choices for the galaxy emission properties. The ones of used here are
those labeled S5 (`late' reionization case) and L20 (`early' reionization
case) which employ an emission spectrum typical of Pop~III stars, a
Salpeter Initial Mass Function (IMF) and an escape fraction of
ionizing photons $f_{esc}=5\%$ and a Larson IMF with $f_{esc}=20\%$
respectively. For details and discussion on the choice of parameters
we refer to CSW and CFWS. These simulations provide the redshift
evolution and the spatial distribution of ionized IGM and have been
used to model the free-free emission from the IGM.

For the diffuse radio emission associated with cluster radio halos and
relics we have used a different simulation of the same cosmological
model, in which in addition to dark matter and baryonic gas, magnetic
fields and cosmic-rays (CR) were also followed. In the simulation, the
cosmological code of Ryu et al. (1993) is used in combination with the
cosmic-ray code {\tt COSMOCR} fully described in Miniati (2001, 2002).
Basically, during the simulation large scale structure shocks are
identified and their jump conditions computed (Miniati et al. 2000).
A fraction of the particles crossing the shock is assumed to be
injected in the diffusive shock acceleration mechanism, and thereafter
accelerated to a power-law spectrum in momentum space. The log-slope
of the distribution is determined by the shock Mach number (${\cal
  M}$) in accord with the test particle limit (e.g. Bell 1978).  The
fraction of injected CR protons is computed according to a variant of
the thermal leakage prescription (e.g. Kang \& Jones 1995); for
practical purposes it amounts to about several $\times 10^{-5}$.  The
fraction of injected CR electrons is assumed to be 100 times smaller
than that (to be roughly consistent with measurements of this
parameter from supernova remnants observations and with EGRET
$\gamma$-ray upper limits at 100~MeV for nearby clusters of galaxies;
Miniati 2002). Both populations of shock accelerated CRs ({\it
  primaries}) are then passively advected with the baryonic flow. In
addition, we follow the evolution of their momentum spectrum by
accounting for all relevant mechanisms of energy loss, such as Coulomb
collisions, adiabatic, bremsstrahlung, synchrotron and inverse Compton
losses for the electrons; and Coulomb and inelastic p-p
collisions for the CR protons. As a byproduct of inelastic p-p
collisions {\it secondary} $e^\pm$ are also generated. Thus with the
same prescription as for the primary electrons, the simulation follows
the evolution of secondaries as well. The simulation employs a
computational box of 50 $h^{-1}$ Mpc on a side, a grid of 512$^3$
cells and 256$^3$ dark matter particles.  With the cosmological
parameters described at the beginning of this Section, each numerical
cell measures about 100 $h^{-1}$ kpc (comoving) and each dark matter
particle corresponds to $2\times 10^9$ $h^{-1}$ M$_\odot$. Finally,
momentum space for the CR components is subdivided into a few
equidistant logarithmic momentum bins as detailed in Miniati (2002).

\section{The 21cm radiation from the intergalactic medium}
The 21 cm hyperfine transition of neutral hydrogen in the IGM provides
a powerful probe for studies of the history of reionization. The physics
of this transition has been well studied in the cosmological context,
initially using linear theory (Scott \& Rees 1990; Kumar et al. 1995;
Madau, Meiksin \& Rees 1997; Tozzi et al. 2000) and more recently
using numerical simulations of reionization (CM; Furlanetto, Sokasian
\& Hernquist 2004). In this work we use the results from the
simulations by CM (which we refer the reader to for further details)
to compare the foreground to the intrinsic 21 cm signal.

The emission in the 21cm line is governed by the spin temperature,
$T_S$. In the presence of the CMB radiation, $T_S$ quickly reaches
thermal equilibrium with the CMB temperature, $T_{\rm CMB}$, and a
mechanism is required that decouples the two temperatures. While the
spin-exchange collisions between hydrogen atoms proceed at a rate that
is too small for realistic IGM densities, Ly$\alpha$ pumping
contributes significantly by mixing the hyperfine levels of neutral
hydrogen in its ground state via intermediate transitions to the $2p$
state. When the first objects form, Ly$\alpha$ pumping will
efficiently decouple $T_S$ from $T_{\rm CMB}$ if $J_\alpha >
10^{-21}\,$ergs cm$^{-2}$ s$^{-1}$ Hz$^{-1}$ sr$^{-1}$. CM find that
the expected diffuse flux of Ly$\alpha$ photons produced by the same
sources responsible for the IGM reionization, satisfies the above
requirement during the `grey age', from redshift $\sim 20$ to the time
of complete reionization.  As the IGM can be easily preheated by
primordial sources of radiation (e.g.~Madau, Meiksin \& Rees 1997;
Chen \& Miralda-Escud\'e 2003), the universe will, most likely, be
observable in 21cm emission at a level that is independent of the
exact value of $T_S$. Variations in the density of neutral hydrogen
(due to either inhomogeneities in the gas density or different
ionized fraction) should appear as fluctuations of the sky brightness
of this transition (e.g, as a brightness temperature increment with
respect to the CMB 
at an observed frequency $\nu$ corresponding to a redshift $1+z=
\nu_o/\nu$, given by $\delta T(\nu)=(T_S - T_{CMB}) \tau
/(1+z)$ where $\tau$ is the optical depth of a patch of IGM in the
hyperfine transition; see e.g. Field 1959). In principle, high
resolution observations of the 21cm line transition in both frequency
(hence redshift) and angle should provide a detailed map of
reionization.

CM studied the evolution of brightness temperature fluctuations in both
the `late' and `early' reionization models (the S5 and the L20 models
respectively, described in Section 2). The S5 model predicts a peak in
the amplitude of the surface brightness fluctuations at $\nu\sim
115$~MHz whereas in the L20 the peak occurs at $\nu \sim 90$~MHz. The
overall amplitude of the signal at its peak, in both models is
$\langle \delta T_{rms}^2 \rangle^{1/2} \sim 10-20$~mK at angular scales $\theta
\sim 5$ arcmin (see CM for details). When comparing the foreground
fluctuations to that of the primary redshifted 21cm we will mainly use
the S5 model where the peak in the 21cm signal occurs at frequencies
which will most easily be probed by the upcoming telescopes (this
choice will not affect any of the analysis performed here). We will
examine the frequency and angular dependence of 21cm signal and
foregrounds in the Conclusions.

\section{Extragalactic Foreground Sources}
In this Section we describe the properties of the sources responsible
for the extragalactic foreground signal. In particular, we describe
how we model their redshift distribution to calculate their spatial
correlation functions and luminosity functions from the simulations.
We will show that our models are able to reproduce reasonably well the
observed populations and hence should provide us with a physically
grounded framework for examining the contribution to the temperature
fluctuations due to unresolved sources.

We consider two main classes of sources.
(a) Point sources: (i) radio galaxies/AGN which produce
synchrotron emission in their nuclei; (ii) HII regions inside ionizing 
halos which produce free-free emission. (b) Extended sources: 
(i) HII regions in the IGM which produce free-free emission;
(ii) cluster radio halos and (iii) cluster radio relics, which
both produce synchrotron emission. As we shall see in 
Section~\ref{acff.se}, radio
galaxies, cluster radio halos and relics appear to provide the most
significant contribution to the sky brightness temperature
fluctuations at angular scales of a few arcmin.

\subsection{Radio Galaxies}
In this Section we describe the simple scheme we have implemented in
the simulations to reproduce the properties of the observed radio
galaxy population. The physical processes that determine the amount
of radio power from active galactic nuclei (AGN) are poorly understood.
However, the radio power in AGN must be linked to the presence of the
supermassive black hole in center of galaxies and to the accretion of
gas onto it.  Several groups have discussed the link between
cosmological evolution of AGN and the formation history of galaxies
using semianalytic models of galaxy formation (see e.g.;~Monaco et
al.~2000; Kauffmann \& Haehnelt~2000; Cavaliere \& Vittorini 2000;
Wyithe \& Loeb~2002; Volonteri, Haardt \& Madau 2003; Haiman, Ciotti
\& Ostriker 2004 and references therein) and recently using
cosmological hydrodynamic simulations (Di Matteo et al. 2003ab). Such
models have been shown to successfully reproduce the B-band luminosity
functions of QSOs and in some cases the X-ray luminosity functions as
well. Here we make the hypothesis that the radio activity is also
linked to the growth and accretion of gas onto black holes and hence
to the activity seen in the optical and X-ray bands. In particular, we
use a prescription similar to that adopted in previous works to couple
the cosmological simulations described in Section 2 with a simple
scheme for the radio population. Note that we shall not attempt to
build a detailed model for the radio galaxy population but merely
reproduce its main observed properties such as the luminosity
functions and their angular clustering.
\begin{figure}
\psfig{figure=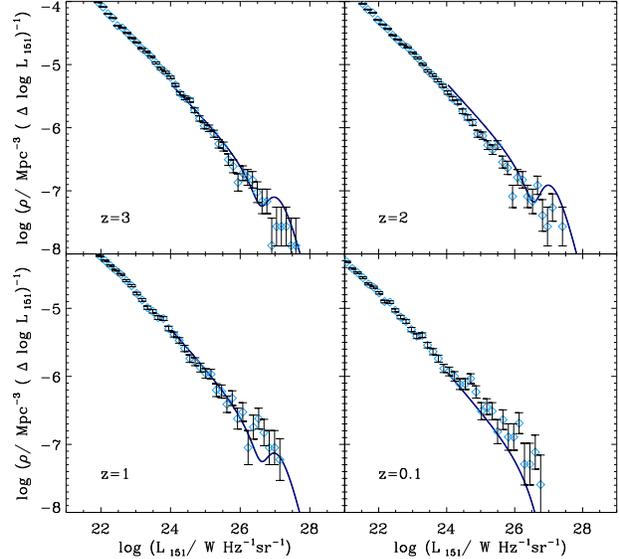,height=8cm}
\caption{\label{fig1}\footnotesize{The radio luminosity function 
at $150$ MHz
in the simulations (diamond symbols) plotted with Poisson error
bars compared to the analytical fits (solid lines) of the observed 
luminosity functions from Willott et al.~(2001) from the 3CRR,
6CE and 7CRS samples.}} 
\end{figure}

The fraction of gas accreted by the central black hole in galaxies
must be related to the observed correlation between stellar velocity
dispersion and black hole mass (Ferrarese \& Merritt 2000; Gebhardt
et al. 2000). Using the most recently revised set of black hole mass
and velocity dispersion measurements (Tremaine et al. 2002), the
$M_{\rm BH}-\sigma$ relation gives: 
\beq M_{\rm BH}=(1.5 \pm 0.2)
\times 10^{8} \Msun \left(\frac{\sigma}{200 \kmps}\right)^{4},
\label{eqn_Mbh_sigmaobs}
\eeq
where $\sigma$ is the velocity dispersion of stars in the bulge.
Ferrarese (2002) and Baes et al. (2003)
find that the stellar
velocity dispersion measured in galaxies is strictly correlated with
the asymptotic value of the circular velocity, $V_{c}$: 
\beq
\log V_{c} = 0.9 \log\sigma + 0.3.
\eeq 
A $V_{c}-\sigma$ relation consistent with the observations has also 
been measured from the simulations (Eq.~8; Di Matteo et al. 2003a).
In order to avoid introducing any details on how  gas cooling,
star formation and feedback is linked to black hole fueling,
we simply combine the previous two relations to find the mass
accreted by a supermassive black hole. We assume this mass to 
be accreted at the Eddington rate after a timescale $t_{Q} \sim
t_{dyn} \sim 0.1 r_{vir}/V_{c}$, where $t_{dyn}$ is the dynamical
timescale estimated at a tenth of the virial radius, $ r_{vir}$.
All galaxies are assumed to undergo active phases with a duty cycle
given simply by $f_{Q} =t_{Q}/t_{H(z)}$ where $t_{H(z)}$ is the Hubble
time at redshift $z$. This fraction
gives us the expected number density of active black holes at 
a given redshift.
\begin{figure}
\psfig{figure=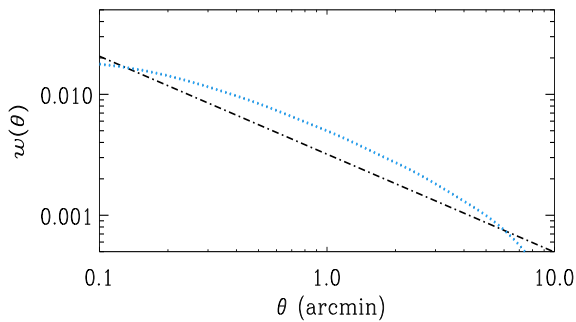,height=5cm,width=8cm}
\caption{\label{fig2}\footnotesize{Angular correlation function
of the simulated radio galaxies (dotted line) compared
to a power law with slope (in log-log) $1-\gamma = 0.8$ and
amplitude $A=2\times 10^{-3}$. The amplitude and shape of the
angular correlation function is consistent with the one
inferred from observations of
faint radio galaxies in the NVSS and FIRST surveys.}}
\end{figure}
We then use the empirical relation found by Merloni, Heinz \& Di
Matteo (2003) and Falcke, Koerding \& Markoff (2003) to
infer the radio luminosity. For each galaxy of a given
nuclear X-ray luminosity and black hole mass, the relation gives,
\beq
\log L^{RG}_{\rm
  R}= 0.6 \log L_{\rm X} +0.8 \log M_{BH} + 7,
\eeq
where $L_{\rm X}$ is the X-ray luminosity in the $2-10$ keV band,
which is related to the bolometric luminosity with a correction $
L_{\rm X} \sim 0.01 L_{\rm Bol}$ (Elvis et al. 1994; Elvis, Risaliti
\& Zamorani 2002). 
In Eq.~3, $L^{RG}_{\rm R}$ is the radio luminosity
at 5~GHz. 
In Fig.~\ref{fig1}, we show the radio luminosity
functions at $150$ MHz and for $z=3,2,1,0.1$ obtained from the
simulations (diamond symbols) using $L^{RG}_{R}$ in Eq.~3 and assuming a
power law spectrum which gives us the luminosity of the radio sources
at a given frequency simply as $L_{\nu}^{RG} \propto \nu^{-\alpha}$
with spectral index $\alpha = 0.8$.  The results from the simulations
are compared with the luminosity functions at $150$ MHz most recently
measured by Willott et al. (2001) from the 3CRR, 6CE and 7CRS radio
surveys (shown with the solid lines)\footnote{Note that the bumps seen
  in the fit to the luminosity function of Willott et al. (2001) are probably
  not real and may be due to restricted number of parameters in their
  model. The important feature is a break in the high and low
  luminosity populations with the low luminosity population
  disappearing at low redshifts.} consistent with the models of Dunlop
\& Peacock (1990). 
Note that because of the limited size of our
simulation box our luminosity function does include a small number of
 bright (very massive) objects.
 
 By projecting the maps from the simulations we also calculate the
 angular correlation function, $w(\theta)$ of the radio galaxy
 population (shown in Fig.~\ref{fig2}) defined as,
\begin{equation}
w(\theta)=\langle \delta({\bf \underline \phi})
\delta({\bf \underline \phi} + {\bf \underline \theta}) \rangle,
\end{equation}
where ${\bf \underline \phi}$, ${\bf \underline \theta}$ are angular
coordinates, $\theta=|{\underline \theta}|$, and $\delta({\bf
  \underline \phi})= [n({\bf \underline\phi})/\langle n \rangle]-1$,
where $n({\bf \underline \phi})$ is number density of galaxies per
unit solid angle at angular position ${\bf \underline \phi}$.  Our
estimate of $w(\theta)$, is calculated from the Limber projection of the
spatial correlation function of radio galaxies. The details of this
calculation are described in Section~5.4. The dash-dotted line in
Fig.~2 shows the conventional power-law form $w(\theta)= A
\theta^{1-\gamma}$ (e.g. Peebles 1993) where $1-\gamma = 0.8$ and $A =
2\times 10^{-3}$. The amplitude $A$ and the shape of the angular
correlation function are fully consistent with the values measured for
cosmological clustering of the 1.4 GHz NVSS and FIRST radio surveys in
the milliJankys and sub-milliJankys populations which show that
$A_{obs}$ is fairly constant at these flux limits and $A_{obs} \sim
1-4 \times 10^{-3}$ for $1-\gamma =0.8$ (Magliocchetti et al.  1998;
Richards 2000; Overzier et al. 2003; Willman et al. 2003). In summary,
our simple model of the radio galaxy population broadly reproduces the
observed luminosity functions and angular clustering of the faint
radio source population. 

\subsection{Free-Free Emission from HII Regions - ISM and IGM}
In this Section we will discuss the free-free emission from
ionized regions, both in the ISM and IGM and how they are
derived from the simulations. 
The free-free luminosity at frequency $\nu$ from an ionized region of 
proper volume $V_{ion}$ is given by $L_\nu^{ff}=\epsilon_\nu V_{ion}$, where
$\epsilon_\nu$ is the free-free emissivity for an hydrogen plasma
at temperature $T$ and electron number density $n_e$, $\epsilon_\nu=
3.2 \times 10^{-39} n_e^2 \; (T/10^4 {\rm K})^{-0.35}$ erg s$^{-1}$ cm$^{-3}$
Hz$^{-1}$ (e.g. Rybicki \& Lightman 1979; Oh 1999). 

The free-free luminosity associated with the ionized ISM of a galaxy
can be estimated by taking the free-free emissivity to be
directly proportional to the recombination rate. In particular,
following  the same prescription used by Oh (1999), we can write the
recombination rate as:
\begin{equation}
\dot{N}_{rec}=\alpha_B n_e^2 V_{ion} \approx (1-f_{esc}) \dot{N}_{ion},
\end{equation}
where $\alpha_B$ is the case B recombination coefficient 
(appropriate for the densities of interest here), $f_{esc}$
is the escape fraction of ionizing photons and $\dot{N}_{ion}$ is
the ionization rate of the galaxy. Using the above equation to
estimate the product $n_e^2 V_{ion}$ and assuming that the temperature
of the ionized regions is $T\sim 10^4$~K, it is
straightforward to derive the free-free luminosity of each galaxy as:
\begin{equation}
L_\nu^{ff,ISM}=3.2 \times 10^{-39} \frac{(1-f_{esc})}{\alpha_B}
\dot{N}_{ion} \; {\rm erg \; s}^{-1} {\rm Hz}^{-1}.
\end{equation}
Once the galaxy emission properties and the value of $f_{esc}$ are set
as described in Section~2, the value of $L_\nu^{ff,ISM}$ is easily
determined for every object. As the mode of star formation is directly
related to the gas metallicity, we expect a change from massive
Pop~III to more standard Pop~II stars once the metallicity is $Z \ge
10^{-4} Z_\odot$ (e.g.; Bromm et al. 2001; Schneider et al. 2001). This
would result in a lower efficiency for production of ionizing photons.
To roughly take this into account we assume that for $z<z_{ion}$,
where $z_{ion}\sim 8$ (13) for the S5 (L20) case, the spectrum
switches to a more standard one, typical of Pop~II stars.

The emissivity from the more diffuse IGM HII regions can be directly
derived from the reionization maps described in Section~2. From these,
the value of $n_e$ and the corresponding emissivity
$\epsilon_{\nu}^{ff,IGM}$ are known for each cell of the simulation
box. After reionization, $\epsilon_{\nu}^{ff,IGM}$ can be derived
assuming that the ionization fraction in each cell is unity.

\subsection{Galaxy Cluster Radio Halos and Relics}
Diffuse, nonthermal radio emission extending over Mpc scales is a well
established observational feature for a fair fraction of galaxy
clusters.  
Diffuse radio sources are classified as either {\it radio halos} when
their morphology is regular and typically centered on and resembling
the X-ray emissivity; or as {\it radio relics} when they are irregular
and located at the periphery of the cluster (in this case the
radiation is typically polarized). In general the volume integrated
emissivity is well characterized by a steep spectrum, $j(\nu)\propto
\nu^{-\alpha}$ with $\alpha \geq 1$ (see e.g. Feretti 2002 for a review). 
\begin{figure}
\psfig{figure=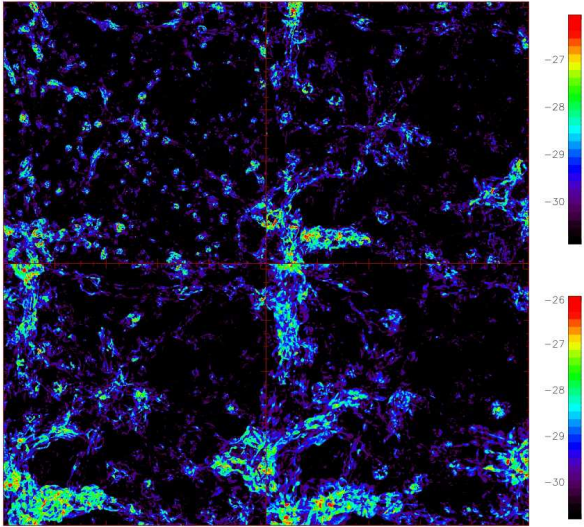,width=8.7cm}
\caption{\label{maps}
\footnotesize{Maps of the 
synchrotron emission (at 150~MHz and in units of erg~s$^{-1}$~cm$^{-2}$
~Hz$^{-1}$~arcmin$^{-2}$) from shock accelerated CR electrons, i.e. 
so called radio relics at redshift $z$=3 (top left), 1 (top right),
0.5 (bottom left) and 0 (bottom left).
The map was obtained by adding the projections of the volume emissivity 
along the three coordinate axes. Thus, it is 50 $h^{-1}$ Mpc on a side 
and 150 $h^{-1}$ Mpc deep.}}
\end{figure}

Statistically it is found that roughly $30-40$\% of rich galaxy
clusters (X-ray luminosity $L_X \geq 10^{45}$ erg s$^{-1}$) do host a
radio halo. In addition, it appears that the radio power (e.g. at 1.4
GHz) of radio halos scales steeply with the virial temperature of the
associated clusters, i.e. $L_{\rm 1.4 GHz} \propto T^{\beta}$ with
$\beta \geq 4$ (Liang et al. 2000). 
The fraction drops below 10\%
when considering smaller clusters (i.e. $L_X \leq 10^{45}$ erg
s$^{-1}$). It is generally believed, although still an open question,
that this is mainly due to the limited sensitivity of
current radio telescope 
(see, e.g., Liang et al. 2000; Bacchi et al. 2003)  

The simulations described in Section 2 (Miniati et al. 2001)
successfully reproduce most of the observed properties of radio halos
and radio relics, particularly those relevant for the purpose of this
paper. Thus, diffuse radio emission produced by synchrotron radiation
from secondary e$^\pm$ is characterized by a morphology and a
low frequency radiation
spectrum consistent with those of radio halos. In addition, the radio
power (at 1.4GHz) scales with the cluster virial temperature with an
index $\beta \sim 4.2$ which is consistent with the observations.
Similarly, synchrotron emission from shock accelerated (primary)
electrons is characterized by a spectrum, morphology and location that
resemble closely those of radio relics.

For CR electrons with a distribution function
$N(\gamma)$, such that $N(\gamma)d\gamma$ is 
the number of particles with Lorentz factor
$\gamma$ between $\gamma$ and $\gamma+d\gamma$,
the synchrotron emissivity
(in units of erg cm$^{-3}$ s$^{-1}$ Hz$^{-1}$)
is computed according to the expression (e.g. Rybicki \& Lightman 1979)
\begin{equation} \label{jsync.eq}
\epsilon_{\nu}^{syn} = \frac{\sqrt{3} e^3 B}{m_ec^2} \sin\alpha
\;
\int_{\gamma_{1}} ^{\gamma_{2}}
F\left(\frac{2\nu\gamma^{-2}}{3\nu_B \sin\alpha} \right) N(\gamma) 
\; 
d\gamma, 
\end{equation}
where $B$ is the magnetic field strength, 
$\alpha$ is the particles pitch angle (over which the $\epsilon_\nu^{syn}$
is further averaged), $\nu_B = eB/(2\pi m_e c)$, is the electron 
nonrelativistic gyrofrequency, $F(x)\equiv x\int_x^\infty dy K_{5/3}(y)$
with $K_{5/3}$ the modified Bessel function of order 5/3,
and all other symbols have their usual
meaning. 

For the diffuse cluster radio halo emission, the population of 
CR electrons, $N(\gamma)$, is provided by secondary e$^\pm$.
Consistently with
current observational constraints, we consider that only 1/3 of
galaxy clusters have radio halos, and further assume that radio halo
formation is suppressed for objects of virial temperature below 1
keV. These two assumptions give us a fairly conservative estimate for
this component.

For cluster radio relics, on the other hand, the emitting electrons,
$N(\gamma)$, are those directly accelerated at cosmic shocks. Even
though observations are sparser for radio relics than radio halos, we
find that the radio power of the simulated relics is consistent with
the values inferred from observations for a broad range of cluster
temperatures (Miniati et al. 2001).  In Fig.~\ref{maps} we show four
slices through the simulations of the emission from radio relics.  We
choose to show the maps for this component in particular to illustrate
the complexity of its morphology. This will be important when
discussing (in Sections 5.3 and 5.4) foreground removal.

\section{Angular clustering of foreground fluctuations} \label{acff.se}

As we shall show in Section \ref{angcl.se}, the angular
brightness temperature fluctuations due to foreground sources
depend on two main quantities: the volume-average comoving 
emissivity and their projected two point correlation function. 
We will start by discussing these quantities for each of the foreground
sources described above.

\subsection{Comoving Emissivities}
For the extended sources and point sources respectively
the volume average comoving emissivity is simply given by, 
\begin{eqnarray}
\langle \epsilon_\nu(z) \rangle&=& 
\frac{1}{V} \left.\int_{extended} \epsilon_\nu({\bf r},z) dV
\right |_{L_{\nu} \le L_{\nu} (S_c,z)} \nonumber \\
&=&\left.\frac{1}{V}\sum \limits_{point} L_\nu(z)\right |_{L_{\nu} \le
  L_{\nu}(S_c,z)}
\label{eq:emi}
\end{eqnarray}
where 
$V$ is the appropriate comoving simulation volume, 
$\epsilon_\nu({\bf r},z)$ in Eq.~\ref{eq:emi} 
is the emissivity at a given point in space, ${\bf r}$, and
$L_\nu$ is the source luminosity which, for extended
sources, is defined as $L_{\nu} = \int \epsilon_{\nu} dV_{eff}$
with the integral performed over the effective volume 
of the extended source (e.g. the cluster).
The integration (summation) is performed over all points
(sources) up to an emissivity (luminosity) corresponding to 
$L_{\nu} \le L_{\nu}(S_c,z) \equiv  S_{c}\times 4 \pi d_{L}^2(z)$
where $S_c$ is the flux cut off above which sources can be resolved
and removed and $d_L$ the luminosity distance 
(cf. Section 5.3).

\begin{figure}
  \vskip -1truecm \psfig{figure=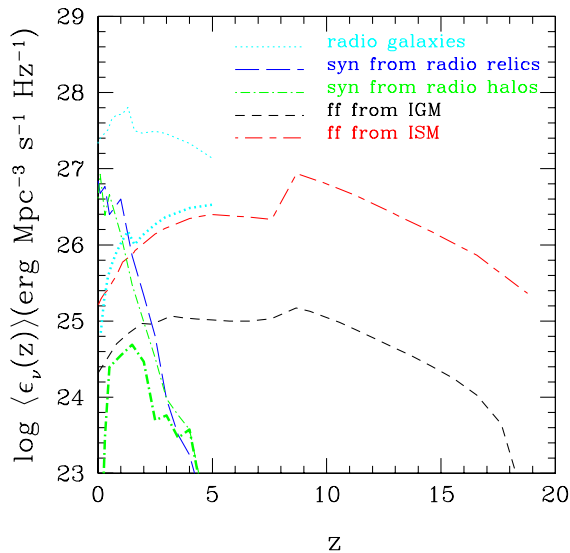,height=9.5cm}
\caption{\label{fig4}\footnotesize{Redshift evolution of the mean comoving 
    emissivity, at $\nu=115$~MHz, due to: free-free from ISM HII
    regions (long-short dashed lines); free-free from IGM HII regions
    (short dashed lines) both in the S5 simulation; radio galaxies
    (thin dotted line and thick dotted line for $S_c=\infty$ and
    $S_c=0.1$ mJy respectively); synchrotron from virialization shocks
    in radio relics (long dashed line); synchrotron from secondary
    electrons in radio halos generated in cosmic rays (thin and thick
    dash-dotted line for $S_c=\infty$ and $S_c=0.1$ mJy
    respectively).}}
\label{emi}
\end{figure}

In Fig.~\ref{fig4} we show the mean emissivity per comoving volume at
the observed frequency $\nu=115$~MHz as a function of redshift for
radio galaxies (dotted line), ISM free-free emission (S5 run;
long-short dashed line), IGM free-free emission (S5 run; short dashed
line), cluster radio relics (long dashed line) and cluster radio halos
(dashed-dotted line) as obtained from the simulations. The thicker
dotted and dash dotted lines show the emissivity when sources above
$S_c = 0.1$ mJy have been removed. Here we will describe the total
emissivities from the various components and reserve the discussion of
source removal to Section 5.3.

As shown in Fig.~\ref{emi}, the radio galaxy, radio halos and relics
were modeled in the simulation out to redshift $z=5$, where most of
the contribution is expected to come from. Below $z=5$, the radio
galaxy comoving emissivity dominates all the other components. Its
peak, at $z\sim 2$, is attributed to the bright component of the
luminosity function which is strongest around this redshift. However,
the overall evolution of the radio galaxy $\langle \epsilon_\nu(z)
\rangle$ is fairly flat. In contrast, for cluster radio halos and
relics $\langle \epsilon_\nu(z) \rangle$ has a much steeper growth, by
a factor $\gsim 10^4$, from $z=5$ to $z=0$. This is due to the rapid
increase in the energy dissipated in shocks during merging/accretion
events as a result of nonlinear breaking of larger wave perturbations
toward lower redshifts (Miniati et al. 2000).
\begin{figure}
\hspace{-1.1cm}
\psfig{figure=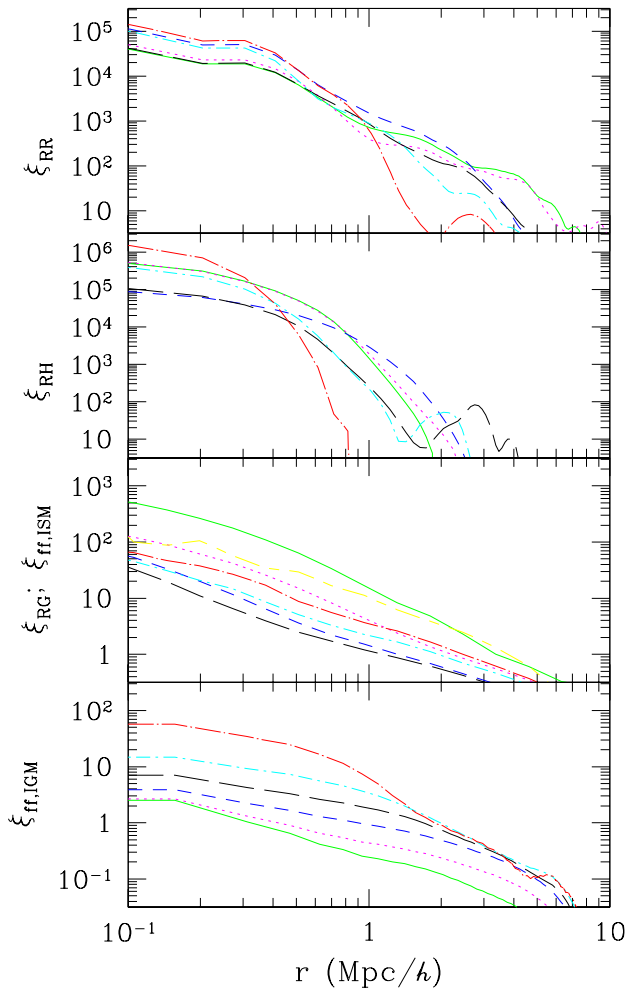,height=14.0cm}
\caption{\footnotesize{The spatial correlation
    functions $\xi(r,z)$ for cluster radio relics, halos, radio
    galaxies (and ISM free-free emission)
    and IGM free-free emission from top to bottom panel
    respectively. The spatial correlation functions are shown at
    $z=0.1,0.5,1,2,3,5$ (solid line, dotted, short dashed, long
    dashed, dot-short dashed, dot-long dashed) for radio halos and
    relics, at $z= 0,1,3,5,10,13,19$ (solid line, dotted, short
    dashed, long dashed, dot-short dashed, dot-long dashed) for the
    radio sources (and ISM free-free emission), 
    and at $z= 13,14,15,16,17,18$ for the free-free emission
    from the IGM (from bottom to top lines).}}
\label{fig5}
\end{figure}
Finally, the free-free emission from ionized regions is significant
only at high redshift, when the other contributions become negligible.
The emissivity from IGM HII regions increases at $z<20$, as ionization
proceeds. Once it is completed, at $z\sim 8$ for the S5 case shown
here, the emissivity decreases as the gas density decreases.  The
emissivity from ISM HII regions is instead proportional to the star
formation rate. The drop observed in the curve is related to the drop
in the ionizing photon production rate following reionization, as
described in Section 4.2. This emissivity is consistent with the
estimate of Bianchi, Cristiani \& Kim (2001), which is derived
independently based on the observed star formation rate and a
population synthesis model. In any case, the contribution from ISM HII
regions is always greater than that from IGM HII regions, with the
ratio reaching a maximum of $\sim 700$ at $z \sim 20$, when the IGM is
mainly neutral, and a minimum of a few at lower redshift.  In summary,
we have shown that the radio galaxies, cluster radio relics and halos
provide the largest contributions to the total comoving emissivity at
moderate/low redshifts. The resulting brightness temperature
fluctuations however, are also a function of the spatial clustering of
the various components (described in Section 5.2).

\subsection{Spatial clustering}
We quantify the spatial clustering of the various sources via the
redshift dependent two-point correlation function, $\xi(r,z)$, defined
as (Peebles 1993): 
\beq
\xi(r,z) = \langle \delta({\bf x},z)
\delta({\bf x + r},z) \rangle,
\label{eq:xi}
\eeq
where for extended sources, $\delta({\bf x},z) = \epsilon_\nu({\bf x},
z)/\langle \epsilon_\nu(z)\rangle-1$  and $\epsilon_\nu({\bf x},z)$
is the emissivity at a point with coordinate ${\bf x}$ at
redshift $z$. 
In practice, to evaluate $\xi(r,z)$ we have computed the power spectrum
$P_{\delta}$ of $\delta({\bf x},z)$, and  Fourier transformed:
\beq
\xi(r,z)_{extended}=\frac{1}{(2\pi)^3}\int P_{\delta}(k) \frac{\sin kr}{kr}
4\pi k^2dk.
\eeq
For the point source populations (radio galaxies and free-free
emission from ISM)
$\delta({\bf x},z) = n({\bf x},z)/\langle n(z) \rangle-1$,
where $n({\bf x},z)$ is number density of galaxies
at ${\bf x}$.
We evaluate the correlation function using
the standard pair estimator, 
\beq
\xi(r,z)_{point}= \frac{N(r,z)} {N_{random}(r,z)}-1,
\eeq
where $N(r,z)$ is the number of pairs separated by a distance $r$,
and $N_{random}(r,z)$ is the number of pairs in the same $r$-bin
expected if galaxies were randomly distributed. 
Because our simulated box is not periodic, we generate random catalogues to
compute $N_{random}(r,z)$.
\begin{figure}
\psfig{figure=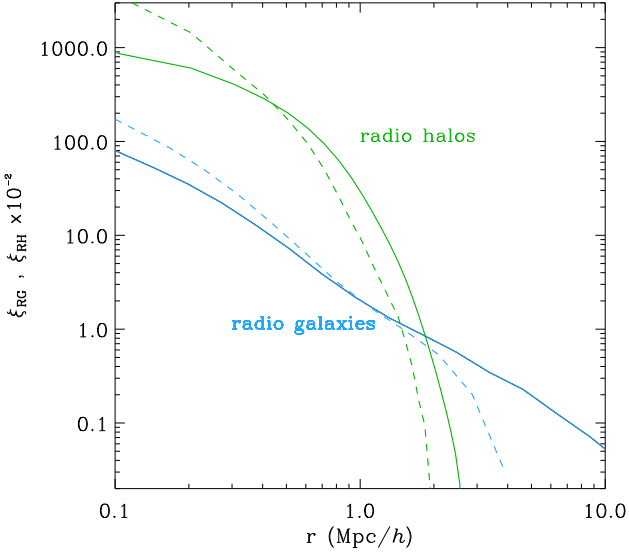,height=8cm}
\caption{\label{fig6}\footnotesize{The spatial correlation
function for radio galaxies and radio halos at $z=2$ and $z=1$ 
respectively. The solid lines show the correlation functions of the 
total sample, the dashes those for sources below the 
flux limit $S_{c} = 0.1 $~mJy. }}
\end{figure}
From top to bottom, Fig.~\ref{fig5} shows the computed $\xi(r,z)$ for
the cluster radio relics , $\xi_{RR}$, halos, $\xi_{RH}$, radio
galaxies , $\xi_{RG}$, (the same as free-free emission from ISM, 
$\xi_{ff, ISM}$ as they are the same halos)
and free-free emission from IGM, $\xi_{ff, IGM}$ (see caption of
Fig.~\ref{fig5} for details). According to our model, the radio
galaxies constitute a random sample of all halos (the duty cycle is
not mass dependent, c.f. Section 4.1). As we expect, the correlation
function of the galaxy population is close to a power law of the form
$\xi(r) \sim (r/r_0)^{-\gamma}$ (usually slightly suppressed on scales
$r \simlt 20-50$ kpc) and with a correlation length, $r_0$ which
varies in the range of $r_0 \sim 2-5h^{-1}$Mpc in the redshift range
studied (Fig.~\ref{fig5}). Note that the amplitude of the correlation
function grows with decreasing redshift for $z \approxlt 10$. At
higher redshift the trend is reversed probably as an effect of the
mass resolution (when halos of mass corresponding to the resolution
limit start representing the high end tail of the mass function). This
is not important as all of the signal from point sources is dominated
by relative low redshifts (c.f.; Section 5.4).

The correlation functions for radio halos and radio relics have
typically large amplitudes, a rather flat inner region and a steep
decline at $r \approxgt 1h^{-1}$Mpc (particularly in the case of radio
halos). The correlation length for cluster radio halos is set
predominantly by the size itself of the brightest sources, which is
roughly of order of a Mpc or so. Radio relics are typically more
extended, as reflected also by their correlation functions, which
decline less steeply than in the case of radio halos. This is a
consequence of the fact that accretion shocks can cover larger regions
than that of the cluster, particularly when two or more
groups/clusters are approaching before a merger (see Fig.~\ref{maps}).

Finally, for the correlation functions of the IGM free-free emission
we only show those corresponding to $z>z_{ion}$.  We note that as
ionization proceeds the HII regions, which initially are isolated and
surrounding the galaxies, will become increasingly more extended
and, as expected, will start overlapping. This has the consequence that
the correlation amplitude decreases with decreasing redshifts.
Because the IGM free-free emissivity shown in Section 5.1 and its
correlation function are negligible with respect to the other
components (and in particular with respect to free-free emission from
the ISM) we will not consider it further.

\subsection{Bright source removal}
As part of the actual experiment, bright enough sources will be
directly identified as foreground contaminants, and hence removed from
the observational maps without contributing to the brightness
temperature fluctuations. The flux limit above which sources can be
removed, $S_{c}$, will of course be dependent on the instrument
sensitivity. Here we take $S_{c} = 0.1$~mJy, corresponding to the
expected sensitivity for source detection by LOFAR and discuss the
major effects that this has on both the emissivities and correlations
functions.

Note that removal of point sources will be far easier than the removal
of extended components. Here we shall illustrate the effects of
applying a flux cut on radio galaxies and on cluster radio halos
(because, as we shall show in Section 5.4, these are the two dominant
contaminants of each class). For cluster radio halos,
we go back to the volume emissivity distribution calculated from the
simulation and calculate the total radio luminosity of each
identified cluster by integrating the synchrotron emissivity over its
volume (roughly out to the virial radius, $r_{vir}$), 
$L_{\nu} = \int_0^{r_{vir}}
\epsilon_{\nu} 4\pi r^2dr$.  The cluster radio emission is then
removed from the volume emissivity distribution if $L_{\nu} \ge
S_{c} \times 4\pi d_L^2$.  As shown by Fig.~\ref{maps} and discussed
in Section 4.3 the morphology of radio relics is fairly complex. We
will not attempt here to remove this extended component. However, we do
expect it to provide similar advantages as those gained from the
removal of cluster radio halos which we describe below. Nevertheless,
such an operation for cluster radio relics may turn out more difficult
due to their more irregular morphology.

As shown in Fig.~\ref{emi}, 
the removal of sources above the detection threshold has the
obvious effect of decreasing the volume emissivity 
(c.f; Eq.~\ref{eq:emi}). Because of the relatively low flux
cut level the emissivity drops dramatically towards the
low redshifts. In particular, for radio galaxies 
emissivity decreases between a factor of $\sim 10$ to $\sim 200$
from $z=5$ to $z=0.1$ 
with respect to the case when no sources have been removed.
For radio halos removal of sources above $S_c =0.1$~mJy
significantly affects the emissivity below $z \sim 2$.
At $z < 2$ the emissivity decreases by up to a factor of a few
thousand.
\begin{figure}
\psfig{figure=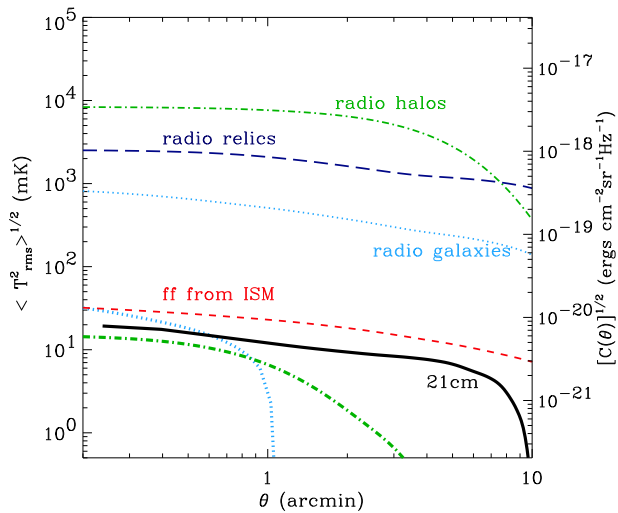,height=7.7cm}
\caption{\label{trms}\footnotesize{Prediction
    for the correlation signal owing to intensity fluctuations of
    radio galaxies (dotted lines), ISM free-free emission (short
    dashed), radio relics (dashed dotted), radio halos (long dashed)
    at $\nu=115$~MHz. The thicker lines show the signal when
    sources above a flux $S_c = 0.1$~mJy are removed. The solid line
    shows the primary correlation signal due to the redshifted 21cm
    emission (CM)}}. 
\end{figure}

In addition to affecting the overall emissivity, the removal of bright
sources has an important effect on the spatial correlation functions. As
an illustrative example, we show in Fig.~\ref{fig6} the correlation
function for radio galaxies and radio halos at $z=2$ and $z=1$
respectively (solid lines) compared with the case in which a flux cut,
$S_c=0.1$~mJy is applied (dashed lines), and all the objects with flux
exceeding $S_c$ have been removed from the respective simulated
catalogues.  Typically, removing the bright sources has the effect of
steepening the spatial correlation function and hence decreasing the
correlation lengths.  In particular, for the radio galaxies, the
amplitude at scales $r \simlt 1 h^{-1}$Mpc is boosted, but at scales
larger than the typical correlation length it is significantly
suppressed. For the case of radio halos, the effect is similar and the
suppression of the correlation functions occurs at even smaller scales
then in radio galaxies. Also, because the correlation functions of
these sources decline very steeply the overall effect is more dramatic
at large scales.

For the radio galaxies, we see from Fig.~\ref{fig6} that the
correlation function of objects below a flux cut is biased in a scale
dependent way with respect to the whole population. This is very
different to the alternate case of galaxies brighter
than flux $S_c$. For the latter, we find that the correlation
function of the brighest galaxies is similar in shape to that of the
whole population (linear biasing), even on scales significantly
smaller than the correlation length (as found by e.g., Mo \& White
1996). Note that we have checked that the effects we have just
described, scale dependent bias versus linear bias, are the same when
a mass cut is adopted instead of a flux cut. We have checked this by
calculating the correlation function $\xi$ for objects below and above
a given dark mass limit $M_{min}$.  This is expected as in our model
we have directly linked e.g., the radio galaxy population to the dark
matter potential. The relative change in clustering of halos after
applying an upper mass threshold is not often considered, but the
boost in clustering on small scales which results, along with
suppression on linear scales, is expected in hierarchical models of
structure formation (e.g. Sheth \& Tormen 1999).

Finally, we note that, as we shall show in the next Section, the
dependence of the spatial clustering on the flux cutoff is important
when the spatial functions are mapped into the angular clustering
signal. Such an effect had not been take into account previously in
calculations of sky intensity fluctuations due to clustering of
foreground sources within the context of the 21cm experiment.

\subsection{Angular Clustering} \label{angcl.se}
\begin{figure*}
\center{
\psfig{figure=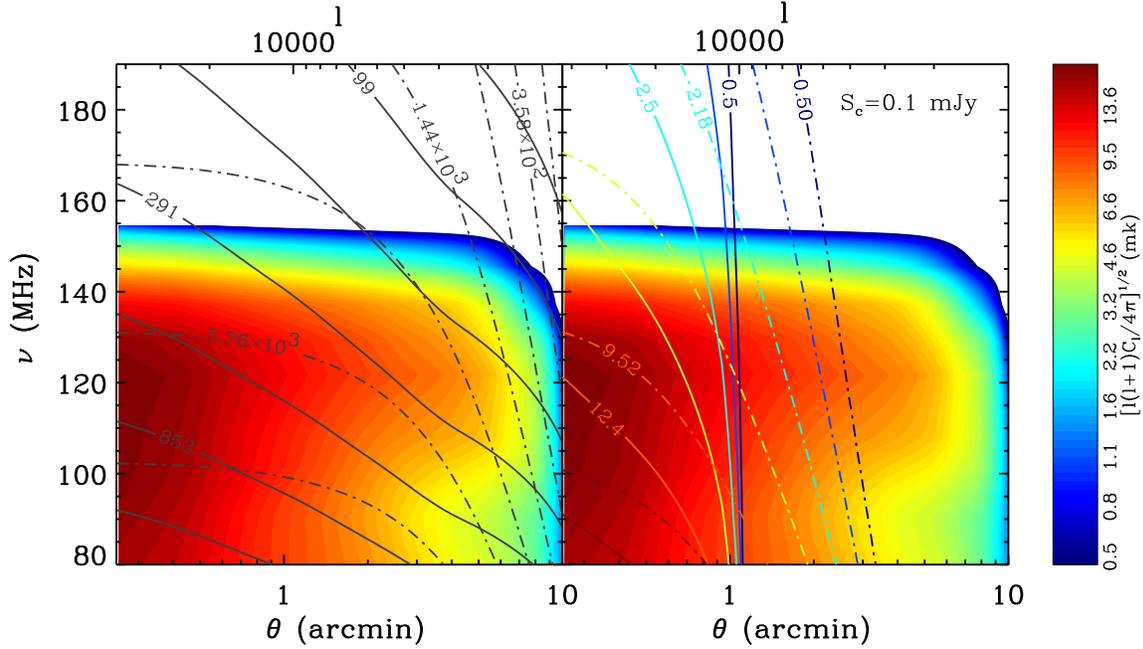,height=10cm}}
\caption{\label{cont}\footnotesize{Contours of 
    the angular power spectrum (in mK) of the the 21cm fluctuations
    (shaded area) and the angular power spectra due to intensity
    fluctuations of radio galaxies (solid contours) and radio halos
    (dot-dashed contours) as a function of angle ($l$) and 
    observed frequency.
    The lines in the right panel show the contours for the total
    foreground signals. In the left panel, foreground sources below a
    flux limit $S_{c} \approxgt 0.1$ mJy have been removed.}}
\label{cont}
\end{figure*}
We now calculate the contribution to the intensity fluctuations,
$\delta I$, owing to the inhomogeneities in the space distribution of 
the various sources described above. The angular
correlation function of such intensity fluctuations is given by,
\beq
C(\theta) = \langle \delta I({\bf \phi})
\delta I({\bf \phi} + {\bf \theta}) \rangle,
\end{equation}
where ${\bf \phi}$, ${\bf \theta}$ are angular coordinates,
and $\delta I({\bf \phi})= [I({\bf \phi})/\langle I
\rangle]-1$, where $I({\bf\phi})$ is the intensity 
at angular position ${\bf \phi}$.

Following the standard approach, we use the Limber projection
(Limber 1953; Peebles 1993)
to calculate, $C(\theta)$ 
(in units of erg$^2$ s$^{-2}$ cm$^{-4}$ Hz$^{-2}$ sr$^{-2}$):
\begin{equation}
C(\theta)=\frac{1}{(4\pi)^2}\int_0^\infty dV_1 dV_2 \frac{\langle{\epsilon}_\nu
(z_1)\rangle}{4\pi d_L^2(z_1)}\frac{\langle{\epsilon}_\nu(z_2)\rangle}
{4\pi d_L^2(z_2)}
\xi(r,z_1,z_2),
\label{eq:limb1}
\end{equation}
where $dV_1, dV_2$,
$\langle\epsilon_{\nu}(z_1)\rangle$, $\langle\epsilon_{\nu}(z_2)\rangle$  
are the comoving volumes and
emissivities at redshifts $z_1$ and $z_2$ respectively. 
For small angles $\theta$ and small radial distances,
$z_1 \sim z_2 \sim z$  and we write:
\beq
C(\theta)=\frac{1}{(4\pi)^2}\int_{z_{min}}^{z_{max}} dz 
\frac{\langle{\epsilon}_\nu(z)\rangle^2}{(1+z)^2} \frac{dx}{dz}
\int_{u_{min}}^{u_{max}} du \; \xi(r,z),
\label{eq:limb2}
\end{equation}
where $\xi(r,z)$ is the spatial two point correlation function,
$r^2=u^2+x^2 \theta^2$, and $x$ is the comoving coordinate, related to
the redshift (for the flat models) by: 
\beq x=\frac{c}{H_0} \int_0^z [\Omega_0(1+z')^3 +
\Omega_{\Lambda}]^{-0.5}{dz'}, 
\eeq 
and we have written $dz_1\; dz_2 =
(dz/dx) \;du \;dz$ for $x_1 \sim x_2 \sim x$ in Eq.~\ref{eq:limb1}.
Analogously, using the standard Limber equation the angular
correlation function, $w(\theta)$ (c.f., Eq.~4), can be written as, 
\beq
w(\theta)=\int_{z_{min}}^{z_{max}}\int_{u_{min}}^{u_{max}}du\; dz
\frac{dz}{dx}\; \xi(r,z).
\end{equation}
We can then simply rewrite the angular clustering of intensity
fluctuations in terms
of a brightness temperature fluctuation in the sky as:
\begin{equation}
\left< T_{rms}^2 \right>^{1/2}=\left [ \frac {l(l+1)C_l}{4 \pi} \right ]
^{1/2} \left (\frac {\partial B_\nu}{\partial T} \right )^{-1},
\end{equation}
where,
\begin{equation}
\frac {\partial B_\nu}{\partial T}=\frac{2k}{c^2} \left( \frac{kT}{h}
\right) ^2 \frac{x^4 {\rm e}^x}{({\rm e}^x-1)^2},
\end{equation}
with $x=h\nu/kT$ and $T=2.725$~K.
The power spectrum, $C_l$ (in units of erg$^2$ s$^{-2}$ cm$^{-4}$ Hz$^{-2}$ 
sr$^{-1}$), is related to $C({\theta})$ by the Legendre transform as:
\begin{equation}
C_l=2\pi\int d({\rm cos}\theta) P_l({\rm cos}\theta) C(\theta)
\sim 2\pi\int_0^\infty d\theta \theta J_0(l\theta) C(\theta),
\end{equation}
where $J_0$ is the Bessel function and the last step is valid for $l
\gg 1$.  Fig.~\ref{trms} shows the prediction for the correlation
signal caused by the intensity fluctuation due to radio galaxies
(dotted lines), ISM free-free emission (short dashed), cluster radio
relics (dashed dotted), cluster radio halos (long dashed) at
$\nu=115$~MHz, the frequency at which the 21cm emission peaks,
according to S5 model of CM. The thinner lines show the total signal
when no flux cutoff has been applied, while the thicker lines show the
signal when sources above a flux $S_c = 0.1$~mJy (corresponding to the
planned detection limit of LOFAR) are removed. The solid line shows
the primary correlation signal due to the redshifted 21cm emission.
The total $T_{\rm rms}$ signal due to any of the foreground components
exceeds the primary 21cm signal. In particular, the extended sources
such as the radio halos and radio relics provide fluctuations about
three (and more) orders of magnitude and the radio galaxies about two
orders of magnitude larger than the 21cm emission through the whole
range of angular scales. The rms fluctuations due to free-free
emission from the ISM, instead, are close to those of the 21cm.
However, the effective brightness temperature fluctuations due to
foreground sources after the removal of sources with $S >S_c$ (for
illustration we have applied the removal only to the highest point and
extended components of the rms fluctuations, e.g. radio halos, and
radio galaxies), decrease significantly. For the radio galaxies,
detection and removal of sources above $S_c = 0.1$ mJy decreases the
fluctuations due to this component down to the level of the 21cm
emission at scales $\theta \lsim 1$ arcmin. Above these angular scales
the foreground signal drops out completely. This is a result of the
combined effects of the decrease of the overall emissivity (e.g.,
Fig.~\ref{emi}) and the suppression of the the amplitude of spatial
correlation of faint radio sources at the scales that map into angular
scales of $\theta \gsim 1$ arcmin (where radial scales and angular
scales are roughly related as, $r \sim 3 (\theta/1') [(1+z)/1]^{0.2}
h^{-1}$ Mpc). For this flux cut, angular scales of $\theta \gsim 1$
arcmin are thus free of this contaminating signal, rendering the 21cm
tomography free of the radio galaxy contamination. A similar result
is obtained when the flux cut is applied to the radio halos. The
decrease in amplitude in this rms fluctuation is even more pronounced
than in the case of radio galaxies. The signal drops to the level of
the primary 21cm at $\theta \approxlt 1$ arcmin and is also well below
it at larger angular scales. Applying the same flux cut to the ISM
free-free emission sources, would lead to a similar effect as for the
radio galaxies, but in this case, the signal would drop well below
that of the 21cm on all scales (and below the axis range plotted in
Fig~\ref{trms}).

After removing sources with flux above $S_c =0.1$~mJy, the dominant
contribution to the rms fluctuations from radio halos comes from $0.6
\approxlt z \approxlt 2$. Similarly for the case of the radio sources
(which however show a different evolution of their mean emissivity and
different clustering), the main contribution to the rms fluctuations
is in the range $0.5 \approxlt z \approxlt 2.5$. This is set by
the the projection of the emissivity and correlation function as
shown in Eq.~\ref{eq:limb2}.

\section{Conclusions}
We have examined the contribution of extragalactic foregrounds to the
angular brightness temperature fluctuations at the frequencies
relevant for observations of the redshifted 21cm signal. We have
used a set of cosmological simulations to model the evolution of radio
galaxies, free-free emission from the ISM and IGM and radio halos and
relics. We calculated their comoving space density emissivity and
spatial correlation functions and projected the simulation volumes to
predict their expected sky temperature fluctuations. 

Because high resolution observations of the 21cm signal will be
carried out in both frequency and angle it is best to summarize the
main conclusion from our analysis by showing the intrinsic and
foreground signals as a function of both scale and frequency. 
Fig.~\ref{cont} shows the
angular power spectrum of the 21cm signal (from CM simulations), and
that of the radio galaxies and radio halos (the two dominating
foreground components for which source removal was studied here) as a
function of frequency (where we have taken the spectral index
of radio galaxies and radio halos to be $\alpha = 0.8$ and 1 
respectively) and angle (scale $l$). In summary,
\begin{itemize}
\item As previously emphasized by Di Matteo et al. (2002, for the case
  of radio galaxies) and Oh \& Mack (2002, for the case of free-free
  emission from the ISM) we have found that the absolute foreground
  signal fluctuates more strongly than that of the 21 cm signal at all
  angular scales and frequencies (left panel Fig.~\ref{cont}). In
  addition, here we have shown that the largest fraction of these
  fluctuations is due to extended cluster radio halos and relics, both
  of which had previously been neglected in the context of the 21cm
  tomography experiment (see however Waxman \& Loeb 2000). Radio
  emission from the cores of galaxies contributes more strongly than
  the free-free emission from their ISM. The presence of foreground
  signal many orders of magnitude above the primary signal needs to be
  addressed because it may seriously hinder the detection of the 21 cm
  line emission as the latter could be easily mimicked by small
  errors/irregularities in the substraction of the foreground spectra.
  
\item We have considered source removal above a flux limit
  $S_c=0.1$~mJy, (close to LOFAR sensitivity) for point sources and
  radio halos (as an example of an extended source). We have found
  that, as expected, the amplitude of the foreground power spectra of
  both radio halo and radio galaxies decreases at all scales. In
  particular, with the additional spatial information on the
  foreground sources provided by the simulations, we have shown that
  the amplitude of the angular foreground fluctuations drops well
  below that of the 21cm at scales $\theta \approxgt 1$ arcmin and
  becomes comparable to it at scales smaller that $\theta \sim 1$ arcmin
  (right panel Fig.~\ref{cont}). At angular scales larger than
  $\theta \sim 1$ arcmin and at frequencies $\nu \approxlt 150$ MHz
  (simply corresponding to the redshifts prior to full reionization in
  the S5 model) the 21cm signal should be detected directly. At scales
  smaller than $\theta \sim 1$ arcmin, substraction of the foregrounds can be
  feasibly carried out in frequency space, as the overall foreground
  signal is at most of the same order as the primary 21cm. Comparing
  maps closely spaced in frequency, (as discussed in detail by
  Zaldarriaga et al. 2004) after the removal of bright sources should
  make the detection of the signal feasible at small scales also
  (where the foreground power spectra are different than the one of
  the 21 cm; Fig.~\ref{cont})\footnote{We note further that the amplitude of
  the 21cm signal shown in Fig.~\ref{cont} corresponds to a given
  bandwidth $\Delta\nu=\nu\Delta z/(1+z)=1$MHz and is expected to
  increase with bandwidth resolution. CM show that the
  signal can increase by a factor up to 4 at small angles and high
  frequencies if e.g.; $\Delta\nu=0.1$MHz can be achieved making
  foreground substraction less crucial at scales $\theta \approxlt 1'$}.
\end{itemize}

From Fig.~\ref{cont}, we conclude that bright sources contribute to
the angular power spectrum at $\theta \approxgt 1$ arcmin and, once
removed, the power on these scales is suppressed. This effect is due
to a scale dependent bias of the spatial correlation functions of
sources below a certain flux cut. This result emphasizes the
importance, when determining the angular fluctuation and power spectra
of foregrounds, of using simulations to project the detailed spatial
distribution of foregrounds as a function of flux cut. Analytical
estimates of brightness temperature fluctuations (Di Matteo et al.
2001; Oh \& Mack 2002), which had to assume a certain angular
clustering of the radio foregrounds, appear to be insufficient for
determining the effect of these contaminants on all scales. This had
lead previous authors to generally more pessimistic conclusions for
the feasibility of 21cm tomography. We find instead that, due to the
effects of bright source removal, 21cm observations carried out at
angular scales $\theta$ of the order of a few arcmin should not be
strongly affected by extragalactic foregrounds. However, how source
substraction can be carried out from the observed maps, needs detailed
work. The maps we have simulated will need to be folded with the
instrument response functions as a function of angle and frequency
(and other specific details of the instrument design). This is
currently being investigated for the case of LOFAR by Valdes et al.
(2004).

We note that determining the angular scale at which the foreground
signal should drop out is, of course, subject to some uncertainty
associated to specifics of source removal. Besides needing to account
for the instrument response, which is beyond the scope of this paper,
this will of course be a function of the theoretical approach we have
adopted to model the foreground source populations down to flux limits
currently unobserved. In fact, it is part of the scientific goals of
future high sensitivity low frequency radio facilities such as LOFAR,
PAST and SKA to sample for the first time faint radio galaxies and
diffuse cluster radio emission (radio halos and radio relics, see
En\ss lin \& R\"ottgering 2002, Br\"uggen et al. 2004) and hence
determine the properties of these population up to high redshifts.

\section*{Acknowledgments}
We would like to thank F. Stoehr for providing us with his
simulations, S. Bertone for the fof catalogues, R. Croft, H.
Rottgering, J. Peterson, S. Bianchi, A. Diaferio and M. Magliocchetti
for many useful discussions.  The paper has been partially supported
by the Research and Training Network ``The Physics of the
Intergalactic Medium'' set up by the European Community under the
contract HPRN-CT-2000-00126.

\label{lastpage}

\newpage

\end{document}